\documentclass{article}
\usepackage{amssymb}
\usepackage{latexsym,amssymb}
\usepackage{graphics}
\usepackage{pstricks}
\usepackage{amsmath}
\usepackage{subfigure}
\usepackage{vmargin}
\usepackage{epsfig}

\begin{document}

\large

\begin{titlepage}

\title{\bf Neutrinos in Warped Extra Dimensions}

\vskip 1cm

\author{G. Moreau \footnote{e-mail: {\tt greg@cftp.ist.utl.pt}} ,
J. I. Silva-Marcos \footnote{e-mail: {\tt juca@cftp.ist.utl.pt}} \\ \\
{\it  CFTP, Departamento de F\'{\i}sica} \\
{\it  Instituto Superior T\'ecnico,
Avenida Rovisco Pais, 1} \\
{\it 1049-001 Lisboa, Portugal}}
\maketitle

\vskip 1cm

\begin{abstract} 
Amongst the diverse propositions for extra dimensional scenarios, the
model of Randall and Sundrum (RS), which
offers a solution for the long standing puzzle of the gauge hierarchy
problem, has attracted considerable
attention from both the theoretical and experimental points of view. In
the context of the RS model with gauge
bosons and fermions living in the bulk, a novel type of mechanism has
arisen for interpreting the strong
mass hierarchy of the Standard Model fermions. This purely geometrical
mechanism is based on a
type and flavor dependent localization of fermions along a warped
extra
dimension. Here, we find concrete
realizations of this mechanism, reproducing all the present experimental
data on masses and mixings of the entire
leptonic sector. We consider the case of Dirac neutrino masses (due to an
additional
right handed neutrino) where the various constraints on RS parameter space
are taken into account.
The scenarios, elaborated in this paper,
generate the entire lepton mass hierarchy and mixing, essentially, via the
higher-dimensional mechanism, as the Yukawa coupling dependence is
chosen to be minimal.
In addition, from the above mechanism, we predict the lepton mixing angle
$10^{-5} \lesssim \sin \theta_{13} \lesssim 10^{-1}$,
a neutrino mass spectrum with normal hierarchy
and the smallest neutrino mass to lie in the range: $10^{-11}
\mbox{eV} \lesssim m_{\nu_1}
\lesssim 10^{-2} \mbox{eV}$. A large part of the $\sin\theta_{13}$
interval should be testable in future
neutrino experiments. 
\end{abstract}

\vskip 1cm

PACS numbers: 11.10.Kk, 12.15.Ff, 14.60.Pq, 14.60.St

\end{titlepage}

\section{Introduction}

\label{intro}

At the moment, string theory \cite{String} is the main candidate which
allows to incorporate gravity into a quantum framework unifying the
elementary particle interactions. String theory is based on the existence of
additional spatial dimensions \cite{Kaluza,Klein}. Recently, a renewed
interest for those extra dimensions has arisen due to several original
proposals for universal extra dimension models \cite{UED} (in which all
Standard Model fields may propagate in extra dimensions) as well as brane
universe models \cite{Akama}-\cite{Antoniadis} (in which Standard Model
fields live in our 3-dimensional spatial subspace) or intermediate models 
\cite{IntermA,IntermB,IntermC} (in which only gauge bosons and Higgs fields
propagate in extra dimensions while fermions are `stuck' at fixed points
along these dimensions). In particular, among the brane universe models, the
two different scenarios suggested by Arkani-Hamed, Dimopoulos and Dvali
(ADD) \cite{ADD,AADD,ADD2} (with large flat extra dimensions) and by Randall
and Sundrum (RS) \cite{Gog,RS}\footnote{%
See also \cite{+-+A}-\cite{CryIII} for extensions of the RS model.} (with a
single small warped extra dimension) have received considerable attention.

Extra dimensional models constitute alternatives to the extensively studied
supersymmetric theories \cite{SUSY}, in the sense that these models have the
following advantages. First, the ADD and RS brane models address a long
standing puzzle: the gauge hierarchy problem (huge discrepancy between the
gravitational scale and electroweak scale). Secondly, a unification of gauge
couplings possibly occurs either at high scales ($\sim 10^{16}$ GeV) \cite%
{UNI-RS}-\cite{UNI-bulk} within small warped extra dimension models or at
lowered scales ($\sim $ TeV) \cite{UNI-ADD,UNI-large} within large flat
extra dimension models. Finally, from a cosmological point of view, there is
a viable Kaluza-Klein WIMP candidate \cite{KKrelic} for the dark matter of
the universe in both the universal extra dimension \cite{LKP1,LKP2,LKP3} and
warped geometry \cite{LZP1,LZP2} models.

The additional interest for extra dimensional models concerns the mysterious
origin of the strong mass hierarchy existing among the different generations
and types of Standard Model (SM) fermions. These models have lead to
completely novel types of approach in the interpretation of the SM fermion
mass hierarchy, which is attractive, as it does not rely on the presence of
any new symmetry in the short-distance theory, in contrast with the
conventional Froggatt-Nielsen mechanism \cite{FN} which introduces a `flavor
symmetry'. Indeed, the interpretation is purely geometrical, and is based on
the possibility that SM fermions have different localizations along extra
dimension(s) which depend upon the flavor and type of fermions, a scenario
realizable in both the ADD \cite{AS}-\cite{Nuss} (see \cite{hep-ph/9912265}-%
\cite{ASreal2} for concrete realizations) and RS \cite{RSloc} (see \cite%
{RSmany} for a realization in RS extensions) models. \newline
At this level, one may mention the other higher-dimensional views suggested
in literature \cite{others1}-\cite{others2} for generating an important SM
fermion mass hierarchy. More specifically, some higher-dimensional ideas
have been proposed, within the ADD \cite{VolMesPow1,VolMesPow2,VolMesPow0},
RS \cite{GNeubertA,GNeubertB,GNeubertC} or RS extension \cite{mypaper}
frameworks, in order to explain the lightness of neutrinos relatively to
other SM fermions.

In this paper, we investigate the possibility that the SM fermion mass
hierarchy is created through the type and family dependence of fermion
locations within the warped geometry of the RS model (the RS model does not
need any new energy scale in the fundamental theory, in contrast with the
ADD model). In such a scenario, the quark mass hierarchy as well as the CKM
mixing angles can be nicely accommodated as shown in \cite{HSquark}. Here,
we will construct the specific 
concrete realizations of this scenario in the leptonic
sector. More precisely, we will determine the domains of parameter space
with minimum fine-tunning (describing fermion locations), which reproduce
all the present experimental data on leptonic masses and mixing angles, and
relying only on a minimal dependence of the Yukawa coupling structure. The
domains of parameter space obtained in this way will give rise to
predictions on the neutrino sector.

Within the context of different SM fermion locations along a warped extra
dimension, the case of Majorana neutrino masses has already been studied:
the results are that neutrino masses and mixing angles can be accommodated
in both scenarios where neutrinos acquire masses through dimension five
operators \cite{HSHHLL} and via the see-saw mechanism \cite{HSseesaw}. In
our present article, we consider the case of Dirac neutrino masses within
the minimal scenario where a right handed neutrino is added to the SM
fields. In a preliminary work \cite{HSpre} on Dirac neutrinos, the charged
lepton Yukawa coupling constants were assumed to be diagonal in flavor space
(for reasons of simplification). However, this is equivalent as to introduce
unexplained strong hierarchies among the Yukawa coupling constants. In
contrast, in our present article, we assume the natural quasi universality
of all lepton Yukawa coupling constants so that the lepton mass hierarchy is
solely governed by the above higher-dimensional mechanism. Therefore, in our
framework, the whole lepton mass hierarchical pattern, can be interpreted in
terms of the higher-dimensional mechanism, thus solving the lepton mass
hierarchical problem.

The organization of this paper is as follows. In Section \ref{frame}, we
describe the effective lepton mass matrices arising when the leptons possess
different localizations along the warped extra dimension of the RS model. In
Section \ref{const}, we make a short review of the experimental constraints
applying on the considered RS scenario. In Section \ref{real}, we
concentrate on the phenomenological implications of the model and present
the domains of parameter space that reproduce the last experimental set of
data concerning the whole leptonic sector. Then, in Section \ref{pred}, some
predictions on the neutrino masses and mixing angles are given and compared
with the sensitivities of future neutrino experiments. Finally, we conclude
in Section \ref{conclu}.

\section{Theoretical framework}

\label{frame}

\subsection{The RS geometrical configuration}

\label{RSgeom}

The RS scenario consists of a 5-dimensional theory in which the extra
spatial dimension (parameterized by $y$) is compactified on a $S^{1}/\mathbb{%
Z}_{2}$ orbifold of radius $R_{c}$ (so that $-\pi R_{c}\leq y\leq \pi R_{c}$%
). Gravity propagates in the bulk and the extra spatial dimension is
bordered by two 3-branes with tensions $\Lambda _{(y=0,\pi R_{c})}$ (vacuum
energy densities) tuned such that, 
\begin{equation}
\Lambda _{(y=0)}=-\Lambda _{(y=\pi R_{c})}=-\Lambda /k=24kM_{5}^{3},
\label{RStensions}
\end{equation}%
where $\Lambda $ is the bulk cosmological constant, $M_{5}$ the fundamental
5-dimensional gravity scale and $k$ a characteristic energy scale (see
below). Within this background, there exists a solution to the 5-dimensional
Einstein's equations which respects 4-dimensional Poincar\'{e} invariance.
It corresponds to a zero mode of the graviton localized on the positive
tension brane (namely the 3-brane at $y=0$) and to the following
non-factorizable metric, 
\begin{equation}
ds^{2}=e^{-2\sigma (y)}\eta _{\mu \nu }dx^{\mu }dx^{\nu }+dy^{2},\ %
\mbox{with}\ \sigma (y)=k|y|,  \label{RSmetric}
\end{equation}%
$x^{\mu }\ [\mu =1,\dots ,4]$ being the coordinates for the familiar 4
dimensions and $\eta _{\mu \nu }=diag(-1,1,1,1)$ the 4-dimensional flat
metric. The bulk geometry associated to the metric (\ref{RSmetric}) is a
slice of Anti-de-Sitter ($AdS_{5}$) space with curvature radius $1/k$.

Let us now describe the physical energy scales within this RS set-up. While
on the 3-brane at $y=0$ (referred to as the Planck-brane) the gravity scale
is equal to the (reduced) Planck mass: $M_{Pl}=1/\sqrt{8\pi G_{N}}=2.44\
10^{18}\mbox{GeV}$ ($G_{N}\equiv $ Newton constant), on the other 3-brane at 
$y=\pi R_{c}$ (called the TeV-brane) the gravity scale, 
\begin{equation}
M_{\star }=w\ M_{Pl},  \label{RSratio}
\end{equation}%
is affected by the exponential \textquotedblleft warp\textquotedblright\
factor $w=e^{-\pi kR_{c}}$. From Eq.(\ref{RSratio}), we see that for a small
extra dimension such that $R_{c}\simeq 11/k$ ($k$ is typically of order $%
M_{Pl}$), one has $w\sim 10^{-15}$ so that $M_{\star }=\mathcal{O}(1)%
\mbox{TeV}$. Hence, the gravity scale $M_{\star }$ on the TeV-brane can be
of the same order of magnitude as the electroweak symmetry breaking scale.
Moreover, if the SM Higgs boson is confined on the TeV-brane, it feels a
cut-off at $M_{\star }=\mathcal{O}(1)\mbox{TeV}$ which guarantees the
stability of Higgs mass with respect to divergent quantum corrections.
Therefore, the RS model completely addresses the gauge hierarchy question. 
\newline
Besides, by considering the fluctuations of the metric (\ref{RSmetric}), one
obtains (after integration over $y$) the expression for the effective
4-dimensional gravity scale as a function of the three fundamental RS
parameters ($k$, $R_{c}$ and $M_{5}$): 
\begin{equation}
M_{Pl}^{2}=\frac{M_{5}^{3}}{k}(1-e^{-2\pi kR_{c}}).  \label{RSkrelat}
\end{equation}%
The feature that the effective 4-dimensional gravity scale is equal to the
high Planck mass $M_{Pl}$ insures that gravitational interactions appear to
be weak from the 4-dimensional point of view, according to experience.

\subsection{The effective lepton mass matrices}

\label{effectiveM}

In order to generate the SM fermion mass hierarchy through the considered
higher-dimensional mechanism, the fermions must reside in the bulk \footnote{%
The behavior of fermions in the bulk was investigated in \cite{GNeubertA}.}
(this is also required for the existence of a Kaluza-Klein WIMP candidate in
the RS model). Then, the SM gauge bosons must live in the bulk as well, if
the 5-dimensional gauge invariance is to be maintained \footnote{%
The consequences of SM gauge bosons in the bulk were studied in \cite%
{HRizzo,Pom} and in \cite{hep-ph/9912498,hep-th/9912232} the complete SM was
put in the bulk.} (gauge bosons have to be in the bulk also, to permit the
gauge coupling unification in the RS model). \newline
Another condition necessary to produce the SM fermion mass hierarchy,
already mentioned in Section \ref{intro}, is that the SM (zero mode)
fermions have different localizations along the warped extra dimension of
the RS model. For that purpose, each type of SM fermion field $\Psi _{i}$ ($%
i=\{1,2,3\}$ being the flavor index) is coupled to its own 5-dimensional
mass $m_{i}$ in the fundamental theory as, 
\begin{equation}
\int d^{4}x\int dy\ \sqrt{G}\ m_{i}\bar{\Psi}_{i}\Psi _{i},  \label{mass}
\end{equation}%
where $G=e^{-8\sigma (y)}$ ($\sigma (y)$ is defined by Eq.(\ref{RSmetric}))
is the determinant of the RS metric. In order to modify the localization of
zero mode fermions, the masses $m_{i}$ must have a non-trivial dependence on
the fifth dimension, more precisely, a `(multi-)kink' profile \cite%
{Rubakov,Rebbi}. The masses $m_{i}$ could be the Vacuum Expectation Values
(VEV) of some scalar fields. An attractive possibility is to parameterize
the masses as \cite{Tamvakis}, 
\begin{equation}
m_{i}=c_{i}\ \frac{d\sigma (y)}{dy}=\pm \ c_{i}\ k,  \label{VEV}
\end{equation}%
where the $c_{i}$ are dimensionless parameters. The masses (\ref{VEV}) are
compatible with the $\mathbb{Z}_{2}$ symmetry ($y\rightarrow -y$) of the $%
S^{1}/\mathbb{Z}_{2}$ orbifold. Indeed, they are odd under the $\mathbb{Z}_{2}$
transformation, like the product $\bar{\Psi}_{\pm i}\Psi _{\pm i}$ (as
fermion parity is defined by: $\Psi _{\pm }(-y)=\pm \gamma _{5}\Psi _{\pm
}(y)$), so that the whole term (\ref{mass}) is even. \newline
Taking into account the term in (\ref{mass}), the equation of motion in
curved space-time for a 5-dimensional fermion field, which decomposes as ($n$
labeling the tower of Kaluza-Klein excitations), 
\begin{equation}
\Psi _{i}(x^{\mu },y)=\frac{1}{\sqrt{2\pi R_{c}}}\sum_{n=0}^{\infty }\psi
_{i}^{(n)}(x^{\mu })f_{n}^{i}(y),  \label{dec}
\end{equation}%
admits the following solution for the zero mode wave function along extra
dimension \cite{RSloc,GNeubertA}, 
\begin{equation}
f_{0}^{i}(y)=\frac{e^{(2-c_{i})\sigma (y)}}{N_{0}^{i}},  \label{0mode}
\end{equation}%
where the normalization factor reads as, 
\begin{equation}
N_{0}^{i\ 2}=\frac{e^{2\pi kR_{c}(1/2-c_{i})}-1}{2\pi kR_{c}(1/2-c_{i})}.
\label{Norm}
\end{equation}%
Eq.(\ref{0mode}) shows that, if $c_{i}$ increases (decreases), the zero mode
of fermion is more localized near the boundary at $y=0$ ($y=\pi R_{c}$),
namely the Planck(TeV)-brane.

The SM fermions can acquire Dirac masses via their Yukawa coupling to the
Higgs VEV. This coupling reads as (starting from the 5-dimensional action), 
\begin{equation}
\int d^4x \int dy \ \sqrt{G} \ \bigg ( \alpha_{ij}^{(5)} \ H \bar \Psi_{+ i}
\Psi_{- j} + h.c. \bigg ) = \int d^4x \ M_{ij} \ \bar \psi_{L i}^{(0)}
\psi_{R j}^{(0)} + h.c. + \dots,  \label{Yuk}
\end{equation}
where $\alpha_{ij}^{(5)}$ are the 5-dimensional Yukawa coupling constants,
the dots stand for Kaluza-Klein (KK) excited mode mass terms and the
effective 4-dimensional mass matrix is given by the integral: 
\begin{equation}
M_{ij} = \int dy \ \sqrt{G} \ \frac{\alpha_{ij}^{(5)}}{2 \pi R_c} \ H(y)
f_0^i(y) f_0^j(y).  \label{MassMatrix}
\end{equation}
As discussed in Section \ref{RSgeom}, the Higgs profile must have a shape
peaking at the TeV-brane: we assume the following exponential form, 
\begin{equation}
H(y)=H_0 \ e^{4k(|y|-\pi R_c)},  \label{Hprofile}
\end{equation}
which can be motivated by the equation of motion for a bulk scalar field 
\cite{GWise}. Using the $W^\pm$ boson mass, one can express the amplitude $%
H_0$ in terms of $kR_c$ and the 5-dimensional weak gauge coupling constant $%
g^{(5)}$.

From Eq.(\ref{MassMatrix}), we observe that even for universal Yukawa
coupling constants (here we assume the natural quasi universality: $%
\alpha_{ij}^{(5)}=\kappa_{ij} g^{(5)}$ with $0.9<|\kappa_{ij}|<1.1$,
following the philosophy adopted in \cite{HSquark,HSHHLL}), the SM fermion
mass hierarchy can effectively be created. Indeed, the fermion masses $%
M_{ij} $ can differ greatly (spanning several orders of magnitude) for each
flavor $i,j$ as the overlap between Higgs profile $H(y)$ and zero mode
fermion wave function $f_0^{i,j}(y)$ varies with the flavor. The reason is
that the zero mode fermion wave functions are flavor dependent through their
dependence on $c_i$ parameters (see Eq.(\ref{0mode})).

The analytical expression for the fermion mass matrix (\ref{MassMatrix}),
obtained by integrating over $y$, has been derived in Eq.(\ref{analytic}) of
Appendix \ref{appendA}. This expression involves only the quantities $\kappa
_{ij}$, $kR_{c}$ and $c_{i}$ because the $g^{(5)}$ dependences introduced by
the amplitude $H_{0}$ and Yukawa coupling $\alpha _{ij}^{(5)}$ exactly
compensate each other. Therefore, the dependences of charged lepton and
neutrino Dirac mass matrices of type (\ref{MassMatrix}) read respectively
as, 
\begin{equation}
M_{ij}^{l}=M_{ij}^{l}(\kappa _{ij}^{l},kR_{c},c_{i}^{L},c_{j}^{l})\ \ %
\mbox{and}\ \ M_{ij}^{\nu }=M_{ij}^{\nu }(\kappa _{ij}^{\nu
},kR_{c},c_{i}^{L},c_{j}^{\nu }),  \label{depend}
\end{equation}
where $\kappa _{ij}^{l}$ ($\kappa _{ij}^{\nu }$) are associated to the
charged lepton (neutrino) Yukawa couplings, $c_{j}^{l}$ ($c_{j}^{\nu }$)
parameterize (\textit{c.f.} Eq.(\ref{VEV})) the 5-dimensional masses (%
\textit{c.f.} Eq.(\ref{mass})) for right handed charged leptons (additional
right handed neutrinos) and $c_{i}^{L}$ parameterize the 5-dimensional
masses for fields belonging to lepton $SU(2)_{L}$ doublets (namely both left
handed neutrinos and left handed charged leptons).

\section{Experimental constraints}

\label{const}

\subsection{Large KK masses}

\label{large}

Next, we discuss the different kinds of constraints on the parameters of the
RS model ($k$, $R_{c}$ and $M_{5}$) as well as on the 5-dimensional mass
parameters ($c_{i}^{L,l,\nu }$) within our framework where gauge bosons and
SM fermions reside in the bulk. \newline
\newline
$\bullet $ \textbf{$k$ and $M_{5}$:} The bulk curvature must be small
compared to the higher-dimensional gravity scale ($k<M_{5}$). Thus, the RS
solution for the metric (\textit{c.f.} Eq.(\ref{RSmetric})) can be trusted 
\cite{RS}. In order to consider the most natural case, where there is only
one energy scale value in the RS model, we first assume the limiting
situation (as in \cite{HSquark,HSpre,HSfv}): 
\begin{equation}
k=M_{5}\simeq M_{Pl}.  \label{kA}
\end{equation}%
This equality between $M_{5}$ and $M_{Pl}$ comes from Eq.(\ref{RSkrelat})
together with our choice of the $k$ value and the fact that one must have $%
kR_{c}\simeq 11$, as explained in Section \ref{RSgeom}. \newline
\newline
$\bullet $ \textbf{$R_{c}$:} Furthermore, precision electroweak (EW) data
place constraints on the RS model \cite{EWboundA}-\cite{EWboundE}. The
reason is, that deviations from precision EW observables arise in the
framework of RS model with SM fields (except the Higgs) inside the bulk. We
briefly review these EW constraints. \newline
First, the mixing between the top quark and its KK excited states results in
a new contribution to the $\rho $ parameter which exceeds the bound set by
precision EW measurements \cite{EWboundA,hep-ph/0204002}. Nevertheless,
there is a way to circumvent this problem by choosing a certain localization
configuration for quark fields (or in other words, certain values of the
5-dimensional mass parameters $c_{i}$ for quarks). \newline
Secondly, mixings between the EW gauge bosons and their KK modes (which go
typically like $m_{W}^{2}/m_{KK}^{2}$) also induce deviations from some
precision EW observables, leading to experimental constraints on the RS
model. E.g. considerations regarding modifications of the weak gauge boson
masses lead to the experimental bound: $m_{KK}\gtrsim 10\mbox{TeV}$ \cite%
{EWboundB} where $m_{KK}=m_{KK}^{(1)}(W^{\pm })$ is the mass of first KK
excitation of $W$ gauge boson (the difference between $m_{KK}^{(1)}(W^{\pm
}) $ and $m_{KK}^{(1)}(Z^{0})$ is insignificant in the RS model). The
deviations from W boson coupling to fermions on the Planck-brane (TeV-brane)
constrain the RS model via: $m_{KK}\gtrsim 4\mbox{TeV}$ ($m_{KK}\gtrsim 30%
\mbox{TeV}$) \cite{EWboundB}. This experimental bound depends on the
localization of SM fermion fields in the bulk (and thus on the values of
mass parameters $c_{i}$ for SM fermions) which fixes the effective amplitude
of weak gauge boson coupling to fermions. If the weak gauge boson masses and
couplings are treated simultaneously, one obtains the conservative bound $%
m_{KK}\gtrsim 10\mbox{TeV}$ for a universal value of SM fermion mass
parameters $c_{i}$ lying in the range $[-1,1]$ (and for $10^{-2}<k/M_{5}<1$) 
\cite{EWboundC}. Finally, a global analysis based on a large set of
precision EW observables was performed in \cite{EWboundD} and has yielded a
lower bound on $m_{KK}$ varying typically between $0$ and $20\mbox{TeV}$ for
a universal value of SM fermion mass parameters which verifies: $|c|<1$.

Experimental bounds on Flavor Changing (FC) processes may also constrain the
RS model since significant FC effects can be generated in the RS scenario
with bulk SM fields \cite{HSpre,HSfv,FCbound}, as will be discussed in the
following. \newline
First, the additional exchange of heavy lepton KK excitations prevents the
cancellation (originating from the unitarity of leptonic mixing matrix)
which suppresses the SM contributions to FC processes like lepton decays: $%
\mu \rightarrow e\gamma $, $\tau \rightarrow \mu \gamma $ and $\tau
\rightarrow e\gamma $. For $m_{KK}=10\mbox{TeV}$, one can have some values
of mass parameters $c_{i}^{L,l,\nu }$ reproducing the correct lepton data
(under the hypothesis of Dirac neutrino masses and the assumption of flavor
diagonal charged lepton Yukawa couplings) such that the branching ratios for
these three rare decays (calculated in the RS framework) are well below
their experimental upper limit \cite{HSpre}. \newline
Secondly, the non-universality of neutral current interactions induces
flavor violating couplings, due to the flavor dependence of fermion
localization, when transforming the fields into the mass basis, and thus,
tree-level FC effects are generated through the (KK) gauge boson exchanges.
For $m_{KK}=10\mbox{TeV}$ and certain values of mass parameters $c_{i}^{L,l}$
fitting the known leptonic masses and mixings (if neutrinos acquire Majorana
masses via dimension five operators), all the rates of leptonic processes $%
Z^{0}\rightarrow l_{i},\bar{l}_{j}$, $l_{i}\rightarrow 3l_{j}$, $\mu
N\rightarrow eN$ and $l_{i}\rightarrow l_{j}\gamma $ [$l_{i}=\{e,\mu ,\tau
\} $] (induced by the FC effects mentioned just above) are compatible with
the corresponding experimental bounds \cite{HSfv}. Similarly, for $m_{KK}=10%
\mbox{TeV}$ and some values of the quark parameters $c_{i}$ in agreement
with quark masses and mixings, mass splittings in neutral pseudo-scalar
meson systems can satisfy the associated experimental constraints \cite{HSfv}%
.

In order to take into account the above constraints on the RS model from
precision EW data and results from experimental bounds on FC reactions
(which both depend on the $c_{i}$ parameter values), we fix the first KK
gauge boson mass at the typical value $m_{KK}=10\mbox{TeV}$, because, in the
following, various values of the $c_{i}$ parameters (fitting lepton data)
are considered. This KK mass choice is equivalent\footnote{%
The mass of first gauge boson KK excitation is given by $m_{KK}=2.45\ k\
e^{-\pi kR_{c}}$ in the RS model \cite{EWboundD}.\label{mKK}} (for the $k$
value of Eq.(\ref{kA})) to the value of RS parameter product $kR_{c}$: 
\begin{equation}
kR_{c}=10.83  \label{RA}
\end{equation}%
We stress that the above constraints from precision EW measurements, derived
for universal values of SM fermion parameters $c_{i}$, do not strictly apply
to our analysis, since we will consider flavor and type dependent values of
parameters $c_{i}^{L,l,\nu }$ (so that the whole lepton flavor structure can
be accommodated). Similarly, the above results from FC effect considerations
do not strictly apply to our scenario, because these are obtained for values
of parameters $c_{i}^{L,l,\nu }$ different from the ones we will take here
(and which must fit the Dirac neutrino masses without relying on a strict
Yukawa coupling dependence).

We can check that the $kR_c$ value of Eq.(\ref{RA}) is well consistent with
a resolution of the gauge hierarchy problem (see Section \ref{RSgeom}):
indeed this value leads to a 5-dimensional gravity scale on the TeV-brane of 
$M_\star=4 \mbox{TeV}$ (\textit{c.f.} Eq.(\ref{RSratio})).

Besides, for this choice $m_{KK}=10\mbox{TeV}$, the mixings between the zero
modes of the quarks or leptons and their KK excitations, induced by the
Yukawa couplings (\ref{Yuk}), are not significant, as shown in the studies 
\cite{HSquark,HSHHLL,HSpre,HSfv}. Indeed, the KK fermion excitations are
then decoupled since their masses are larger than (or equal to) $m_{KK}$
(for any $c_{i}$ value) in the RS model \cite{EWboundD}. \newline
The first consequence of this small mixing is that the quark/lepton masses
and mixing angles can be reliably computed from the mass matrix $M_{ij}$ for
the zero modes (see action (\ref{Yuk})) as the mass corrections due to KK
modes can be safely neglected \cite{HSquark,HSHHLL,HSpre,HSfv} (as well as
at the one loop-level \cite{HSquark,Muon}). \newline
Another consequence is that the variation of effective number of neutrinos
contributing to the $Z^{0}$ boson width, induced by mixings between the zero
and the KK modes of neutrinos, is well below its experimental sensitivity,
as shown in \cite{HSpre}, for characteristic values of the parameters $%
c_{i}^{L,\nu }$ (of order unity) reproducing the correct Dirac neutrino
masses and mixing angles. \newline
\newline
$\bullet $ \textbf{$c_{i}^{L,l,\nu }$:} From a theoretical point of view,
the natural values of 5-dimensional masses $m_{i}$ (\textit{c.f.} Eq.(\ref%
{VEV})) appearing in the original action (\ref{mass}) are of the same order
of magnitude as the fundamental scale of the RS model, namely the bulk
gravity scale $M_{5}$, avoiding the introduction of new energy scales in the
theory. Hence, for $k=M_{5}$ (like in Eq.(\ref{kA})), the absolute values of
lepton parameters $c_{i}^{L,l,\nu }$ (defined by Eq.(\ref{VEV})) should be
of the order of unity: 
\begin{equation}
|c_{i}^{L,l,\nu }|\approx 1.  \label{cA}
\end{equation}

Next, we present all the existing bounds concerning the 5-dimensional mass
parameters $c_{i}$. The motivation is to get an idea of what is the typical
range allowed for $c_{i}$ values. Nevertheless, the reader must keep in mind
that these bounds have been obtained under the simplification assumption
that each of the parameters $c_{i}$ are equal to a universal value $c$. This
does not strictly apply to our scenario, where the parameters $%
c_{i}^{L,l,\nu }$ are flavor and type dependent. \newline
For a universal value $c\lesssim 0.3$, considerations on contributions of
virtual KK tower exchanges to fermion pair production (similarly to
contact-like interactions) at colliders (LEP II and Tevatron Run II) force
the $M_{\star }$ value to be significantly in excess of $10\mbox{TeV}$ \cite%
{EWboundD}, disfavoring then the RS model as a solution to the gauge
hierarchy problem. \newline
Bounds can also be placed on $c$ by calculating the contributions to the
anomalous magnetic moment of the muon due to KK excitation exchanges: the
experimental world average measurement for $(g-2)_{\mu }$ translates into
the bound $c\lesssim 0.70$ (for the first KK masses between a few and $10%
\mbox{TeV}$) \cite{Muon}. \newline
Finally, an examination of the perturbativity condition on effective Yukawa
coupling constants yields the constraint $c\lesssim 0.77$ \cite{Muon}.

We end this section with a discussion on the effective couplings of
non-renormalizable four-fermion operators involving lepton fields, as these
depend on the location parameters $c_{i}^{L,l,\nu }$. First, the rare lepton
flavor violating reactions induced by such operators, as for instance the
decay $\mu \rightarrow eee$, are not expected to reach an observable rate,
unless leptons are localized close to the TeV-brane \cite{HSfv}, a
configuration which will never occur for the $c_{i}^{L,l,\nu }$ values that
we consider here. Such operators, for example $Q_{1}Q_{1}Q_{2}L_{3}$ or $%
U_{1}^{c}U_{2}^{c}D_{1}^{c}E_{3}^{c}$, are also dangerous as they permit
proton decay channels \cite{HSquark}. It seems impossible to find quark and
lepton locations which are in agreement simultaneously with the known
fermion masses and the proton life time \cite{HSquark,HSfv}, pointing to an
additional symmetry for example such as baryon or lepton number (protecting
the proton against its instability). A precise analysis of the quark
locations is beyond the scope of our study.

\subsection{Small KK masses}

\label{small}

In this section, we present a different characteristic scenario: we give a
possible set of RS parameters (giving rise to a smaller $m_{KK}$ than the
one mentioned previously) and 5-dimensional mass parameters different from
the ones proposed in the previous section, in agreement with the several
types of constraints, in the case where precision EW constraints are
softened by specific mechanisms (with bulk fermions and gauge bosons). 
\newline
\newline
$\bullet $ \textbf{$R_{c}$:} As in the previous section, we maintain the
parameter product $kR_{c}$ at, 
\begin{equation}
kR_{c}=10.83,  \label{RB}
\end{equation}
so that the TeV-brane gravity scale $M_{\star }=4\mbox{TeV}$, while still
addressing the gauge hierarchy solution. \newline
\newline
$\bullet $ \textbf{$k$ and $M_{5}$:} Precision EW data place a typical bound
on the first KK gauge boson mass $m_{KK}\gtrsim 10\mbox{TeV}$ (see Section %
\ref{large}), which renders the discovery of the gauge boson KK excitations
at LHC (via direct production) quite challenging. Indeed, the LHC (with an
optimistic integrated luminosity of $100fb^{-1}$) will be able to probe $%
m_{KK}$ values only up to about $6\mbox{TeV}$ for a universal absolute value
of the $c_{i}$ parameters smaller than unity \cite{EWboundD}. Nevertheless,
some models have been suggested in order to make the precision EW lower
bounds on $m_{KK}$ less stringent. This we will discuss now. \newline
In \cite{Custodial}, it was proposed to enhance the EW gauge symmetry to $%
SU(2)_{L}\times SU(2)_{R}\times U(1)_{B-L}$, recovering the usual gauge
group via a breaking of $SU(2)_{R}$ on the Planck-brane. The right handed SM
fermions are promoted to $SU(2)_{R}$ doublet fields, with the new (non
physical) component having no zero mode. Hence, for example in the lepton
sector (with an additional right handed neutrino), the right handed $%
c_{i}^{l,\nu }$ parameters would now describe the location of $SU(2)_{R}$
doublets but the total number of $c_{i}^{L,l,\nu }$ parameters would remain
identical. Because of the bulk custodial isospin gauge symmetry arising in
this context, all the precision EW data (including those on the ``oblique''
parameters $S$, $T$ and the shift in coupling of $b_{L}$ to $Z^{0}$) can be
fit with a mass $m_{KK}$ of only a few TeV. We underline the fact that this
result concerning $m_{KK}$ has been obtained for a universal value of $%
c_{i}^{L,l,\nu }$ parameters larger than $1/2$ (in contrast with the non
universal $c_{i}^{L,l,\nu }$ values which will be considered in our
analysis). \newline
Similarly, the brane-localized kinetic terms for fermions \cite{BraneF} 
\textit{or} gauge bosons \cite{BraneB}, which are expected to be present in
any realistic theory (induced radiatively and also possibly present at
tree-level), allow to improve the fit of precision EW observables and to
relax the resulting lower bound on $m_{KK}$ value down to a few TeV (see 
\cite{EWBa,EWBb} for gauge boson kinetic terms and \cite{EWF} for the
fermion case).

Assume that $m_{KK}=1\mbox{TeV}$, and, that one of the above models hold, so
that the precision EW data do not conflict with such a light gauge boson KK
excitation. This $m_{KK}$ value is simultaneously inside the LHC potential
search reach (see above) and compatible with the present collider bound
obtained at Tevatron Run II (with a luminosity of $200pb^{-1}$ and a
center-of-mass energy of $1.96\mbox{TeV}$) on the first KK graviton mass,
namely, $m_{KK}^{(1)}(G)>675\mbox{GeV}$ at $95\%\ C.L.$ (for $k=0.1M_{Pl}$) 
\cite{TevatronA,TevatronB}. Indeed, this bound is equivalent to $m_{KK}>431%
\mbox{GeV}$ since the ratio $m_{KK}^{(1)}(G)/m_{KK}$ is equal to $3.83/2.45$
(\textit{c.f.} foot-note \ref{mKK}) in the RS model \cite{EWboundD}. \newline
For the $kR_{c}$ value given by Eq.(\ref{RB}), the typical mass value $%
m_{KK}=1\mbox{TeV}$ that we have chosen is obtained for (\textit{c.f.}
foot-note \ref{mKK}), 
\begin{equation}
k=0.1M_{Pl}.  \label{kB1}
\end{equation}
Once the $k$ and $R_{c}$ parameter values are known, the $M_{5}$ value is
fixed by Eq.(\ref{RSkrelat}). For the $k$ and $R_{c}$ values corresponding
to Eq.(\ref{RB}) and Eq.(\ref{kB1}), $M_{5}$ is equal to, 
\begin{equation}
M_{5}=1.13\ 10^{18}\mbox{GeV},  \label{kB2}
\end{equation}
Thus, the two values of fundamental energy scales $k$ and $M_{5}$ in the RS
model are quite close.

We note that for the choice $m_{KK}=1\mbox{TeV}$, as in the scenario of
previous section where $m_{KK}=10\mbox{TeV}$, mixings between the zero and
the KK modes of leptons should still not be significant, because the KK
lepton masses (systematically larger than $m_{KK}$) remain typically large
relatively to the zero mode lepton masses. \newline
\newline
$\bullet $ \textbf{$c_{i}^{L,l,\nu }$:} Since the masses (\ref{VEV}) have to
be of the same order as the fundamental scale $M_{5}$, Eq.(\ref{kB1}) and
Eq.(\ref{kB2}) tell us that the natural absolute values of lepton parameters 
$c_{i}^{L,l,\nu }$ read as, 
\begin{equation}
|c_{i}^{L,l,\nu }|\approx 4.6  \label{cB}
\end{equation}

The whole discussions of Section \ref{large}, on all the existing bounds
concerning 5-dimensional mass parameters $c_{i}$ and on the
non-renormalizable operators, still hold within the characteristic framework
considered in this Section.

\section{Realistic RS scenarios}

\label{real}

In this section, we search for the parameter region values which reproduces
all the experimental data on lepton masses and mixing angles.

\subsection{Approximation of lepton mass matrices}

The analysis of the parameter space requires the study of certain limits.
From the formula for lepton mass matrices $M_{ij}^{l,\nu }$ (see Appendix %
\ref{appendA}), it is clear that, in large regions of the parameter space
spanned by $c_{i}^{L}$, $c_{j}^{l,\nu }$, we have to a good approximation: 
\begin{equation}
M_{ij}^{l,\nu }=\kappa _{ij}^{l,\nu }\ g_{_{i}}(c_{i}^{L})\ g_{_{j}}^{\prime}
(c_{j}^{l,\nu })  \label{sep}
\end{equation}%
where the $g_{_{i}},\ g_{_{j}}^{\prime}$ are suitable functions for a
certain region. E.g. for the (as we shall see, important) region $%
1/2<c_{i}^{L}$, $c_{i}^{l}<3/4$, we obtain $g_{_{i}}=\ g_{_{j}}^{\prime}$ $%
=g$, with 
\begin{equation}
g(x)=\sqrt{\frac{m_{0}(kR_{c})(x-1/2)}{4-2x}}\ e^{\pi kR_{c}(2-x)}.
\label{gfunc}
\end{equation}%
This structure of $M_{ij}$ for the lepton mass matrices, has important
consequences and, as we will see in the following, will be helpful for a
clear understanding of the model.

\subsection{The relevant theoretical parameter space}

As mentioned in Eq.(\ref{depend}), the lepton mass matrices depend on the
parameters: $\kappa _{ij}^{l,\nu }$, $kR_{c}$ and $c_{i}^{L,l,\nu }$. 
\newline
\newline
$\bullet $ \textbf{$kR_{c}$:} According to Eq.(\ref{RA}) and Eq.(\ref{RB}),
corresponding to the two considered scenarios of Sections \ref{large} and %
\ref{small}, we take the characteristic value $10.83$ for the parameter
product $kR_{c}$. Nevertheless, our results (for $\kappa _{ij}^{l,\nu }$ and 
$c_{i}^{L,l,\nu }$ values) and predictions (on neutrinos) are not
significantly modified for $kR_{c}$ not exactly equal but only close to $%
10.83$. Different orders of magnitude for $kR_{c}$ are not desirable because
the condition $kR_{c}\approx 11$ is needed for solving the gauge hierarchy
problem. \newline
\newline
$\bullet $ \textbf{$c_{i}^{L,l,\nu }$:} We describe the range for the $%
c_{i}^{L,l,\nu }$ values and our motivations for choosing this range. With
respect to the typical $c_{i}^{L,l,\nu }$ values, the two scenarios proposed
respectively in Sections \ref{large} and \ref{small} are equivalent in the
sense that their characteristic relation (\ref{cA}) and (\ref{cB}) both lead
to $c_{i}^{L,l,\nu }$ with absolute values of order of unity, if one does
not consider high $|c_{i}^{L,l,\nu }|$ values as excluded by the various
existing bounds mentioned at the end of Section \ref{large} (although those
have been deduced under the simplification hypothesis of a unique and
universal $c_{i}^{L,l,\nu }$ value). Motivated by the orders given in Eq.(%
\ref{cA}) and Eq.(\ref{cB}) as well as the existing bounds concerning $%
c_{i}^{L,l,\nu }$ parameters (obtained under the simplification assumption
of a universal $c_{i}^{L,l,\nu }$ value), we restrict ourself to the range: 
\begin{equation}
0.1<|c_{i}^{L,l,\nu }|<5,  \label{searchc}
\end{equation}%
a choice which is appropriate to the two scenarios of large and small $%
m_{KK} $ described in previous section. We notice that by limiting our
analysis to this range, we restrict our search to $c_{i}^{L,l,\nu }$ values,
which generate the wanted lepton mass hierarchy, and which are all of the
same order. The existence of such natural values of the same order, for the
fundamental parameters $c_{i}^{L,l,\nu }$, would confirm the fact that the
strong lepton mass hierarchy can indeed be totally explained by our
higher-dimensional model, in contrast with the SM where Yukawa couplings are
unnaturally spread over several orders of magnitude. \newline
\newline
Some preliminary restrictions on the $c_{i}^{L,l,\nu }$ values may also be
deduced from an analytical study of lepton mass matrices $M_{ij}^{l,\nu }$.
From the trace of squared mass matrix $M_{l}M_{l}^{\dagger }$, which can be
expressed as a function of charged lepton masses: 
\begin{equation}
\sum_{ij}(M_{ij}^{l})^{2}=m_{e}^{2}+m_{\mu }^{2}+m_{\tau }^{2},
\label{trace1}
\end{equation}%
we find that the largest $|M_{ij}^{l}|$, say\footnote{%
Assuming that $|\kappa _{ij}^{l,\nu }|\approx 1$, one can choose $%
|M_{33}^{l}|$ and $|M_{33}^{\nu }|$ to be exactly the largest value, without
imposing any restrictions on the masses and mixings; it is simply a choice
of weak basis.} $|M_{33}^{l}|$, must obey the following relation 
\begin{equation}
\begin{array}{l}
\frac{1}{3}\sqrt{m_{e}^{2}+m_{\mu }^{2}+m_{\tau }^{2}}\leq |M_{33}^{l}|\leq 
\sqrt{m_{e}^{2}+m_{\mu }^{2}+m_{\tau }^{2}}.%
\end{array}
\label{trace2}
\end{equation}%
Thus, each column (or row) of $M_{ij}^{l}$ must have elements which are
large enough to satisfy the relation 
\begin{equation}
\begin{array}{l}
m_{e}m_{\mu }m_{\tau }\ =\ |\det M^{l}|\ \leq \
6(M_{33}^{l})^{2}\sum_{k}|M_{kj}^{l}|%
\end{array}
\label{det1}
\end{equation}%
for any column $(M_{1j}^{l},M_{2j}^{l},M_{3j}^{l})$ (or similarly for any
row). Taking into account that $|M_{kj}^{l}|$ in (\ref{det1}) drops down
very rapidly if $c_{i}^{L}$ or $c_{j}^{l}$ is larger than $1$, we derive
from Eq.(\ref{det1}) the following upper limit: 
\begin{equation}
c_{i}^{L},c_{j}^{l}<1.1  \label{clcr}
\end{equation}%
Assuming hierarchical neutrino masses \footnote{%
In our conventions, the neutrino mass eigenvalues are noted $m_{\nu
_{1,2,3}} $ with $m_{\nu _{1}}<m_{\nu _{2}}<m_{\nu _{3}}$.}, we obtain
similar restrictions on, at least one, of the $c_{j}^{\nu }$, which must not
be too large. One may choose this one to be the $c_{1}^{\nu }$. Using the
relation, $\frac{1}{3}\sqrt{\Delta m_{31}^{2}}$ $\approx $ $\frac{1}{3}\sqrt{%
m_{\nu _{1}}^{2}+m_{\nu _{2}}^{2}+m_{\nu _{3}}^{2}}$ $\leq |M_{33}^{\nu }|$,
we find (with the experimental range for $\Delta m_{31}^{2}$ given in Eq.(%
\ref{4SigmaDataA})), 
\begin{equation}
c_{1}^{\nu }<1.5  \label{crnu}
\end{equation}%
\newline
$\bullet \kappa $\textbf{$_{ij}^{l,\nu }$:} Finally, we discuss the
quantities $\kappa _{ij}^{l,\nu }$ which parameterize the Yukawa couplings
(see end of Section \ref{effectiveM}). We assume that the lepton mass matrices,
and the $\kappa _{ij}^{l,\nu }$, are purely real. In order to reproduce CP
violating observables, one needs to introduce complex phases in the Yukawa
couplings. A comment will be added, at the end of Section (\ref{finalparam}%
), on general complex values. Concerning the absolute value of parameters $%
\kappa _{ij}^{l,\nu }$, we consider the natural range (see discussion at the
end of Section \ref{effectiveM}): 
\begin{equation}
0.9<|\kappa _{ij}^{l,\nu }|<1.1  \label{searchkappa}
\end{equation}%
Indeed, we want to address the question of how much of the phenomenology can
be accommodated, purely, by $kR_{c}$ and $c_{i}^{L,l,\nu }$, the extra
dimensional parameters, thus, reducing the contribution from the SM
parameters $\kappa _{ij}^{l,\nu }$ (proportional to the Yukawa coupling
constants), as much as possible. Therefore, we study the possibility of
obtaining correct masses and mixings in the RS model, for the case $|\kappa
_{ij}^{l,\nu }|=1$, allowing only for small perturbations of this value.
With regard to the signs of the parameters $\kappa _{ij}^{l,\nu }$, let us
first assume that all $\kappa _{ij}^{l,\nu }$ are positive. Just as an
illustrative exercise, suppose all $\kappa _{ij}^{l,\nu }=1$. Then, from the
structure in (\ref{sep}), we obtain for the mass matrices $M_{ij}$ of the
neutrinos and charged leptons 
\begin{equation}
\begin{array}{lll}
\begin{array}{c}
M_{\nu }=D_{L}\cdot \Delta \cdot D_{\nu } \\ 
\\ 
M_{l}=D_{L}\cdot \Delta \cdot D_{l}%
\end{array}
& \ ;\ \ \ \  & \Delta =\left[ 
\begin{array}{ccc}
1 & 1 & 1 \\ 
1 & 1 & 1 \\ 
1 & 1 & 1%
\end{array}%
\right]%
\end{array}
\label{mnch}
\end{equation}%
where $D_{L}=\mathrm{diag}(a_{1,}a_{2},a_{3})_{L}$, $D_{\nu ,l}=\mathrm{diag}%
(b_{1,}b_{2},b_{3})_{\nu ,l}$, where the $a$'s and $b$'s are obtained from
the $g_{_{i}}$ and $g_{_{j}}^{\prime}$ functions in (\ref{sep}). In this\ simple
approximation, only the tau and one neutrino eigenstate have mass.
Furthermore, the resulting squared matrices $H_{\nu }=$ $M_{\nu }M_{\nu
}^{\dagger }$and $H_{l}=$ $M_{l}M_{l}^{\dagger }$ are proportional 
\begin{equation}
\begin{array}{lll}
\begin{array}{c}
H_{\nu }=\rho _{\nu }\ D_{L}\cdot \Delta \cdot D_{L} \\ 
\\ 
H_{l}=\rho _{l}\ D_{L}\cdot \Delta \cdot D_{L}%
\end{array}
& \ ;\ \ \ \  & 
\begin{array}{c}
\rho _{\nu }=\sqrt{b_{\nu _{1}}^{2}+b_{\nu _{2}}^{2}+b_{\nu _{3}}^{2}} \\ 
\\ 
\rho _{l}=\sqrt{b_{l_{1}}^{2}+b_{l_{2}}^{2}+b_{l_{3}}^{2}}%
\end{array}%
\end{array}
\label{hnch}
\end{equation}%
Thus, the matrices $O_{\nu }$ and $O_{l}$, which diagonalize respectively $%
H_{\nu }$ and $H_{l}$, are equal; there is no mixing: $U_{MNS}=O_{l}^{%
\dagger }O_{\nu }={1\>\!\!\!\mathrm{I}}$, and although small deviations from 
$\kappa _{ij}^{l,\nu }=1$ may be sufficient to generate masses for the other
charged leptons and neutrinos, this scenario only leads to small deviations
from $U_{MNS}={1\>\!\!\!\mathrm{I}}$, for the mixing. Therefore, at least
some of the $\kappa _{ij}^{l,\nu }$ must be very different from $\kappa
_{ij}^{l,\nu }=1$, or have different signs. As we maintain $|\kappa
_{ij}^{l,\nu }|$ close to one, we must allow for some $\kappa _{ij}^{l,\nu }$
to be negative, in order to obtain large mixings (and also to obtain some of
neutrino mass differences sufficiently large\footnote{%
In fact, for the neutrinos, at least one of the perturbations of $\kappa
_{ij}^{\nu }$ will have to be as large as $(9/2)\sqrt{\Delta
m_{21}^{2}/\Delta m_{32}^{2}}$ $>0.4$ to account for the neutrino mass
differences \cite{JucaGus}.}). However, the negative signs must be at
different positions in the mass matrices of charged leptons and neutrinos,
otherwise, for similar reasons as explained in (\ref{hnch}), the solar
mixing would be too small. In Appendix \ref{appendB}, based on an analytical
approach, we provide explicit examples of $\kappa _{ij}^{l,\nu }$ sign
configurations which are shown to be satisfactory from the experimental data
point of view.

\subsection{The relevant experimental lepton data}

\label{fitdata}Strictly speaking the lepton masses given by Eq.(\ref%
{MassMatrix}) and Eq.(\ref{Yuk}), that we consider here, are running masses
at the cutoff energy scale of the effective 4-dimensional theory, which is
in the TeV range (if the gauge hierarchy problem is to be treated). If we
consider lepton masses at this common energy scale, of the order of the
electroweak symmetry breaking scale, we avoid the effects of the flavor
dependent evolution of Yukawa couplings on the lepton mass hierarchy. The
predictions for charged lepton masses, obtained from mass matrix (\ref%
{MassMatrix}), will be fitted with the experimental mass values taken at the
pole \cite{PDG}. In order to take into account the effect of the
renormalization group from the pole mass scale up to the TeV cutoff scale
(considered for theoretical masses), and which is only of a few percents 
\cite{hep-ph/9912265}, we assume an uncertainty of $5\%$ on the measured
charged lepton masses. This uncertainty is in agreement with our philosophy
not to determine the fundamental parameter values with too much high
accuracy. For similar reasons, we consider the experimental data on neutrino
masses and leptonic mixing angles only at the $4\sigma $ level \cite{Valle}.

Next, we present in detail the $4\sigma $ data, on neutrino masses and
leptonic mixings, that will be used in this work. A general three-flavor fit
to the current world's global neutrino data sample has been performed in 
\cite{Valle}. The data sample used in this analysis includes the results
from solar, atmospheric, reactor (KamLAND and CHOOZ) and accelerator (K2K)
experiments. The values for oscillation parameters obtained in this analysis
at the $4\sigma $ level are contained in the intervals: 
\begin{equation*}
6.8\leq \Delta m_{21}^{2}\leq 9.3\ \ \ [10^{-5}\mbox{eV}^{2}],
\end{equation*}
\begin{equation}
1.1\leq \Delta m_{31}^{2}\leq 3.7\ \ \ [10^{-3}\mbox{eV}^{2}],
\label{4SigmaDataA}
\end{equation}
where $\Delta m_{21}^{2}\equiv m_{\nu _{2}}^{2}-m_{\nu _{1}}^{2}$ and $%
\Delta m_{31}^{2}\equiv m_{\nu _{3}}^{2}-m_{\nu _{1}}^{2}$ are the
differences of squared neutrino mass eigenvalues, and, 
\begin{equation*}
0.21\leq \sin ^{2}\theta _{12}\leq 0.41,
\end{equation*}
\begin{equation*}
0.30\leq \sin ^{2}\theta _{23}\leq 0.72,
\end{equation*}
\begin{equation}
\sin ^{2}\theta _{13}\leq 0.073,  \label{4SigmaDataB}
\end{equation}
where $\theta _{12}$, $\theta _{23}$ and $\theta _{13}$ are the three mixing
angles of the convenient form of parameterization for the leptonic mixing
matrix (denoted as $U_{MNS}$) now adopted as standard by the Particle Data
Group \cite{PDG}.

In addition to considerations on the measured lepton mass and mixing values
mentioned above, one has also to impose the current experimental limits on
absolute neutrino mass scales. With regard to our case of Dirac neutrino
masses, the relevant limits are the ones extracted from the tritium beta
decay experiments \cite{107,108,109}, since these are independent of the
nature of neutrino mass (in contrast with bounds from neutrinoless double
beta decay results which apply exclusively on the Majorana mass case). The
data on tritium beta decay provided by the Mainz \cite{108} and Troitsk \cite%
{109} experiments give rise to the following upper bounds at $95\%\ C.L.$, 
\begin{equation*}
m_{\beta }\leq 2.2\ \mbox{eV}\ \ \ \mbox{[Mainz]},
\end{equation*}
\begin{equation}
m_{\beta }\leq 2.5\ \mbox{eV}\ \ \ \mbox{[Troitsk]},  \label{mbetaLIM}
\end{equation}
with the effective mass $m_{\beta }$ defined by, $m_{\beta
}^{2}=\sum_{i=1}^{3}|U_{ei}|^{2}m_{\nu _{i}}^{2}$, where $U_{ei}$ denotes
the leptonic mixing matrix elements and $m_{\nu _{i}}$ the neutrino mass
eigenvalues.

\subsection{The obtained parameter values}

\label{finalparam}

In order to find the domains of parameter space with minimum fine-tunning,
and in agreement with all present experimental data on leptons (described in
Section \ref{fitdata}), we have performed a scan on $|c_{i}^{L,l,\nu }|$
values in the range (\ref{searchc}) with a step of $0.01$ simultaneously
with a scan on $|\kappa _{ij}^{l,\nu }|$ values in the range (\ref%
{searchkappa}) with a step of $0.1$. We have considered both signs for $%
c_{i}^{L,l,\nu }$ values. With respect to the $\kappa _{ij}^{l,\nu }$
quantities, we have considered 10 different sign configurations, which
correspond to all possible signs, relevant for the mixing. It is clear, that
certain sign configurations are equivalent (or even irrelevant) as they can
be obtained from each other by weak basis (permutation) transformations.

We find that the $c_{i}^{L,l,\nu }$ values reproducing the present lepton
masses and mixings (\textit{c.f.} Section \ref{fitdata}) correspond to the
two configurations, 
\begin{equation}
\begin{array}{lll}
c_{1}^{L}=0.50-0.52\ ;\ \  & c_{2}^{L}=0.54-0.56\ ;\ \  & c_{3}^{L}=0.54-0.56
\\ 
c_{1}^{l}=0.65-0.66\ ;\ \  & c_{2}^{l}=0.71-0.73\ ;\ \  & c_{3}^{l}=0.56-0.57
\\ 
c_{1}^{\nu }=1.32-1.35\ ;\ \  & c_{2}^{\nu }=1.34-1.36\ ;\  & c_{3}^{\nu
}=1.32-5%
\end{array}
\label{region}
\end{equation}%
and 
\begin{equation}
\begin{array}{lll}
c_{1}^{L}=0.27-0.30\ ;\ \  & c_{2}^{L}=0.41-0.43\ ;\ \  & c_{3}^{L}=0.49-0.50
\\ 
c_{1}^{l}=0.66-0.67\ ;\ \  & c_{2}^{l}=0.62-0.63\ ;\ \  & c_{3}^{l}=0.70-0.71
\\ 
c_{1}^{\nu }=1.42-1.43\ ;\ \  & c_{2}^{\nu }=1.37-1.38\ ;\  & c_{3}^{\nu
}=1.38-5%
\end{array}
\label{region1}
\end{equation}%
It is clear, that these regions are defined modulo permutations among $%
c_{i}^{L}$, $c_{i}^{l}$ or $c_{i}^{\nu }$, which for obvious reasons, do not
change either the mixings or the masses. These correspond to permutations of
the left handed or right handed fields, which, of course, are irrelevant.
The variations in $c_{i}^{L}$, $c_{j}^{l,\nu }$ shown here, are compatible
with values for $kR_{c}\ \approx 11$. Essentially, we find two permitted
regions for $c_{i}^{L}$ and $c_{j}^{l,\nu }$: one where $c_{i}^{L}\gtrsim
1/2 $ and one where $0.27<c_{i}^{L}<1/2$.

In the following, we give an illustrative example, a complete set of
parameters reproducing the charged lepton masses and present data from
neutrino oscillation experiments (\textit{c.f.} Eq.(\ref{4SigmaDataA}) and
Eq.(\ref{4SigmaDataB})). The $c_{i}^{L,l,\nu }$ values 
\begin{equation}
\begin{array}{lll}
c_{1}^{L}=0.50628\ ;\ \  & c_{2}^{L}=0.55\ ;\ \  & c_{3}^{L}=0.555 \\ 
c_{1}^{l}=0.6584\ ;\ \  & c_{2}^{l}=0.714\ ;\ \  & c_{3}^{l}=0.5676 \\ 
c_{1}^{\nu }=1.345\ ;\ \  & c_{2}^{\nu }=1.34\ ;\  & c_{3}^{\nu }=1.365,%
\end{array}
\label{realistic}
\end{equation}
together with $\kappa _{13}^{l}=-1$, $\kappa _{22}^{l}=\kappa _{13}^{\nu
}=\kappa _{31}^{\nu }=\kappa _{32}^{\nu }=1.1$, $-\kappa _{23}^{\nu }=\kappa
_{33}^{\nu }=0.9$ and all other $\kappa _{ij}^{l,\nu }=1$, lead to the
following leptonic observables, 
\begin{equation}
\begin{array}{lll}
m_{e}=0.51\ \ \mathrm{MeV};\ \  & m_{\mu }=105\ \ \mathrm{MeV}\ ;\ \  & 
m_{\tau }=1.77\ \ \mathrm{GeV}\  \\ 
\Delta m_{21}^{2}=7.9\ \ 10^{-5}\ \mathrm{eV};\ \  & \Delta m_{31}^{2}=2.0\
\ 10^{-3}\ \mathrm{eV}\ ;\ \  &  \\ 
\sin ^{2}(\theta _{12})=0.37\ ;\ \  & \sin ^{2}(\theta _{23})=0.64\ ;\ \  & 
\sin ^{2}(\theta _{13})=0.0033%
\end{array}
\label{realmass}
\end{equation}

One may have CP violation if some of the $|\kappa _{ij}|\approx 1$ are
complex. E.g. in the previous example, if we choose $\kappa _{22}^{\nu }$ to
have the small imaginary part $\kappa _{22}^{\nu }=$ $1.0+0.1i$, while
keeping all other input values identical, we obtain already a large $J=|\mathrm{Im}\
(U_{12}U_{23}U_{22}^{*}U_{13}^{*})|=0.003$ (where $U=U_{MNS}$). The masses
and mixings in (\ref{realmass}) do not change significantly.

\section{Predictions on neutrinos}

\label{pred}

\begin{figure}
\begin{center}
\includegraphics[width=10cm,height=7cm]{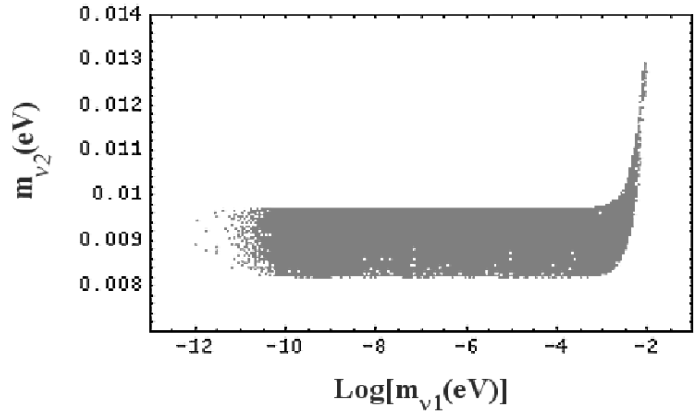}
\end{center}
\caption{Predicted neutrino mass eigenvalues $m_{\protect\nu_1}$ and $m_{%
\protect\nu_2}$ (in eV).}
\label{fig:spectrum}
\end{figure}

\begin{figure}
\begin{center}
\includegraphics[width=10cm,height=7cm]{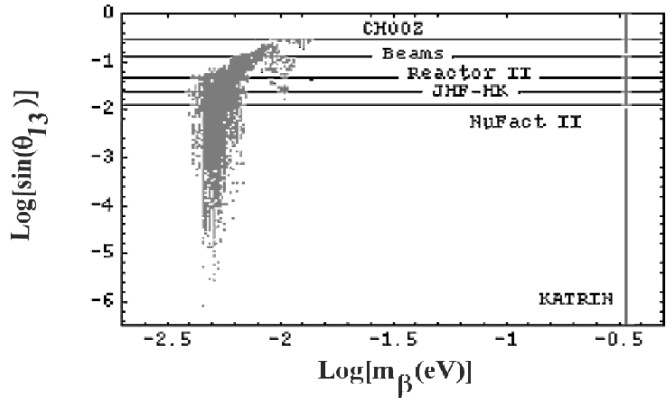}
\end{center}
\caption{Predicted values for $\sin \protect\theta_{13}$ and the effective
mass $m_\protect\protect\beta$ (in eV), as defined in Section \protect\ref%
{fitdata}. We indicate the current CHOOZ bound on $\sin \protect\theta_{13}$
(as issued from the three-flavor global analysis of neutrino data sample 
\protect\cite{Valle} mentioned in Section \protect\ref{fitdata}) as well as
the best sensitivity on $\sin \protect\theta_{13}$ expected for each type of
next coming neutrino experiment (see text), namely the potential reaches for
combined beam projects, JHF-HK, Reactor II and NuFact II. We also mark the
maximum estimated sensitivity on $m_\protect\protect\beta$, planned to be
reached by the next generation tritium beta decay experiment KATRIN, which
is around $0.35 \ \mbox{eV}$ \protect\cite{110}.}
\label{fig:sensitivities}
\end{figure}

We have found the regions of parameter space (see Section \ref{finalparam})
that fit all the experimental values of lepton masses and mixings (\textit{%
c.f.} Section \ref{fitdata}). Those regions of parameter space correspond to
the values of $\sin \theta_{13}$ and neutrino masses shown in Figures (\ref%
{fig:spectrum}) and (\ref{fig:sensitivities}).

First, we comment Fig.(\ref{fig:spectrum}). Except in the region where $%
m_{\nu _{1}}$ gets close to $10^{-2}\mbox{eV}$, $m_{\nu _{1}}$ is negligible
compared to $m_{\nu _{2}}$ (so that $m_{\nu _{2}}\simeq \sqrt{\Delta
m_{21}^{2}}$). The lower and upper limits on $m_{\nu _{2}}$, appearing
clearly in the figure, are given, to a good approximation, by the squared
roots of experimental limits on $\Delta m_{21}^{2}$ (\textit{c.f.} Eq.(\ref%
{4SigmaDataA})). This means that, in this large region, $m_{\nu _{2}}$ takes
nearly all the possible values allowed by present experimental measurements,
or in other words, that the models, and parameter space obtained here, do
not yield particular predictions on $m_{\nu _{2}}$. As $m_{\nu _{1}}$
increases up to $\sim 10^{-2}\mbox{eV}$, $m_{\nu _{2}}$ also increases,
thus, the difference $\Delta m_{21}^{2}$ remains well inside the allowed
experimental range (\ref{4SigmaDataA}). Similarly, this region is not really
predictive for $m_{\nu _{2}}$. A similar conclusion (of low predictability)
also holds for $m_{\nu _{3}}$, which lies typically in the range: $%
[0.03,0.06]\ \mbox{eV}$, and for $m_{\nu _{1}}$, which spans several orders
of magnitude as exhibits the figure\footnote{%
For $c_{3}^{\nu }$ values larger than $5$ (see Eq.(\ref{searchc}) and Eq.(%
\ref{region})-(\ref{region1})), $m_{\nu _{1}}$ would remain in the same
interval as the one exhibited by Fig.(\ref{fig:spectrum}).}.

In Fig.(\ref{fig:sensitivities}), we have plotted the physical quantities
(predicted by our model) that will be measured by the future neutrino
experiments, namely the leptonic mixing angle $\theta _{13}$ and the
effective mass $m_{\beta }$ (defined in Section \ref{fitdata}). \newline
The sensitivity limits on $\sin ^{2}2\theta _{13}$ at $90\%$ \textit{%
confidence level} \cite{hep-ph/0204352,hep-ph/0403068} (see also \cite%
{hep-ph/0303232,hep-ph/0409028}) of future neutrino experiments, designed in
particular at probing this leptonic mixing, are the following ones (for a
normal neutrino mass hierarchy and best-fit values of the other oscillation
parameters); for combined conventional beam experiments: $0.061$ [MINOS,
plus, CNGS experiments ICARUS and OPERA], for superbeams: $0.024$ [NuMI], $%
0.023$ [JPARC-SK], $0.018$ [JHF-SK, a first generation long-baseline
project] and $0.0021$ [JHF-HK, a second generation long-baseline beam], for
reactors: $0.032$ [Double-Chooz] and $0.009$ [Reactor II, a set-up for
projects like KASKA, Diablo Canyon, Braidwood,\dots ] and for neutrino
factories \cite{NeutFact}: $0.0017$ [NuFact I, low luminosity] and $0.00059$
[NuFact II, high luminosity]. \newline
Therefore, this figure shows that the next measurements of $\sin \theta
_{13} $ will only partially allow us to test the studied higher-dimensional
mechanism. This is because $\sin \theta _{13}$ reaches values which are
smaller than the best experimental sensitivities expected. This figure also
demonstrates that future tritium beta decay experiments should not be able
to test the $m_{\beta }$ values predicted by our higher-dimensional
mechanism, which are too low.

So far we have only considered the normal hierarchy for the neutrino mass
spectrum. An important consequence of the structure of mass matrix (\ref{sep}%
), for neutrinos, is that it is impossible to obtain degenerate, almost
degenerate, or even inverse hierarchical neutrinos, unless there is some
precise conspiracy between parameters $c_{i}^{L}$, $c_{j}^{\nu }$ and the
Yukawa couplings $\kappa _{ij}^{\nu }$. E.g. taking $\kappa _{ij}^{\nu }=1$,
clearly, leads to strict hierarchical neutrinos. The same applies if there
is just one $(-)$ sign among the $\kappa _{ij}^{\nu }$ signs (see Appendix %
\ref{appendB})\footnote{%
For one $(-)$ sign, one may have in principle an inverted hierarchy mass
spectrum for neutrinos, but as with two or three $(-)$ signs, this requires
fine-tunning between $g_{_{i}}(c_{i}^{L})$, $\ g^{\prime}_{_{j}}(c_{j}^{\nu})$ and $%
\kappa _{ij}^{\nu}$.}. Two or three crucial $(-)$ signs, i.e. $(-)$ signs
which cannot be eliminated by rephasing the lepton fields, may lead to 3
neutrinos with masses of the same order of magnitude, but to obtain
degenerate neutrinos, one must have fine-tunning between the functions $%
g_{_{i}}(c_{i}^{L})$, $\ g^{\prime}_{_{j}}(c_{j}^{\nu })$ and parameters $\kappa
_{ij}^{\nu }$.

\section{Conclusion}

\label{conclu}

The RS model, with SM fields in the bulk and the Higgs boson on the
TeV-brane, has been studied. We have considered the typical values of the
fundamental RS parameters ($k$, $R_{c}$ and $M_{5}$) which are compatible
with all the relevant constraints: the theoretical constraints, e.g. the
condition required to solve the gauge hierarchy problem, and the complete
set of experimental bounds (from collider data, FC physics, precision EW
measurements,\dots ).

We have found configurations of lepton locations, along the extra dimension,
reproducing all the present experimental data on leptonic masses and mixing
angles, in the case where neutrinos acquire Dirac masses (with a right
handed neutrino added to the SM fields). The neutrino and charged lepton
sectors have been treated simultaneously since these two sectors are related
phenomenologically (via the lepton mixing matrix $U_{MNS}$) and
theoretically (because the left handed neutrinos and charged leptons are
localized in an identical way, since they belong to the same $SU(2)_{L}$
doublet). The field location configurations, we obtained, generate the
entire hierarchy among lepton masses, including the structure in flavor
space, as well as, the lightness of neutrinos relatively to the charged
lepton masses and electroweak scale (providing, in this sense, an
alternative to the usual see-saw mechanism).

Besides, we have determined the domains of parameter space with minimum
fine-tunning, in agreement with all the experimental values of leptonic
masses and mixings. Then, we have deduced, from the domains, predictions on
the two measurable leptonic quantities (with the exception of the physical
complex phases) which are still unknown (more precisely, weakly
constrained), namely the lepton mixing angle $\theta _{13}$ and the ground
of neutrino mass spectrum (only the two neutrino mass squared differences
are fixed by the present experimental results). We predict an approximate
order of magnitude for $\sin \theta _{13}$ lying between $10^{-1}$ and $%
10^{-5}$. A part of this range should fall into the sensitivity interval
reachable by next generation of neutrino experiments, which means that the
studied mechanism, based on the localization of bulk fermions within the RS
model, should be partially testable (from the parameter space point of view)
by future experiments. We also predict a neutrino mass spectrum with the
normal hierarchy and the smallest mass eigenvalue in the interval: $10^{-11}%
\mbox{eV}\lesssim m_{\nu _{1}}\lesssim 10^{-2}\mbox{eV}$. Hence, the studied
model within the RS framework is not especially predictive on neutrino
masses, compared e.g. to an equivalent model (also producing SM fermion mass
hierarchies from flavor and nature dependent locations of fields) in the ADD
framework which predicts: $m_{\nu _{1}}\sim 10^{-2}\mbox{eV}$ (for Majorana
neutrino masses) \cite{ASreal2}.

Our important quantitative result is that the whole lepton mass hierarchy
can be completely explained by the studied higher-dimensional mechanism,
without requiring any special pattern for the Yukawa coupling constants.

\vspace{1cm}

\noindent \textbf{\Large Acknowledgments}

\noindent The authors are grateful to G.~C.~Branco and M.~N.~Rebelo for
useful conversations. G.~M. acknowledges support from a \textit{Marie Curie}
Intra-European Fellowships (under contract MEIF-CT-2004-514138) within the
6th European Community Framework Program.

\newpage

\appendix
\noindent \textbf{\Large Appendix} \vspace{0.5cm}

\renewcommand{\thesubsection}{A.\arabic{subsection}} \renewcommand{%
\theequation}{A.\arabic{equation}} \setcounter{subsection}{0} %
\setcounter{equation}{0}

\section{Fermion mass matrix}

\label{appendA}

Within the studied higher-dimensional scenario where SM fermions possess
various localizations along the warped extra dimension of the RS model, the
effective 4-dimensional fermion Dirac mass matrix is given by, 
\begin{equation}
M_{ij}= \ \kappa_{ij} \ m_0(kR_c) \ \frac{\sqrt{1/2-c_i} \ \sqrt{1/2-c_j}}{%
4-c_i-c_j} \ \frac{e^{\pi k R_c(4-c_i-c_j)}-1}{(e^{2 \pi k
R_c(1/2-c_i)}-1)^{1/2} \ (e^{2 \pi k R_c(1/2-c_j)}-1)^{1/2}}.
\label{analytic}
\end{equation}
\newline
This formula is obtained after integration of expression (\ref{MassMatrix})
over $y$ (on the range $[-\pi R_c,\pi R_c]$), by using Eq.(\ref{0mode})-(\ref%
{Norm}) and Eq.(\ref{Hprofile}). For instance, with the value $kR_c=10.83$
considered in our analysis, the quantity $m_0(kR_c)$ entering Eq.(\ref%
{analytic}) reads as $m_0(10.83)=7.61 \ 10^{-33}\mbox{eV}$.

\renewcommand{\thesubsection}{B.\arabic{subsection}} \renewcommand{%
\theequation}{B.\arabic{equation}} \setcounter{subsection}{0} %
\setcounter{equation}{0}

\section{Sign configurations}

\label{appendB}

In this appendix, we present explicit examples of $\kappa _{ij}^{l,\nu }$
sign configurations which give rise to significant lepton mixings. \newline
\newline
$\bullet $ Let us first assume that all $\kappa _{ij}^{\nu }=\kappa
_{ij}^{l}=1$ except for $\kappa _{33}^{\nu }=\kappa _{33}^{l}=-1$. Then from
(\ref{sep}), we have 
\begin{equation}
\begin{array}{ccc}
\begin{array}{c}
M_{\nu }=D_{L}\cdot \Delta ^{\prime }\cdot D_{\nu } \\ 
\\ 
M_{l}=D_{L}\cdot \Delta ^{\prime }\cdot D_{l}%
\end{array}
& \ ;\ \ \ \  & \Delta ^{\prime }=\left[ 
\begin{array}{ccc}
1 & 1 & 1 \\ 
1 & 1 & 1 \\ 
1 & 1 & -1%
\end{array}
\right]%
\end{array}
\label{mnch1}
\end{equation}
The $(-)$ sign in the $33$ position of the mass matrices $M_{\nu }$ and $%
M_{l}$ has two important consequences. First, it leads to a non zero mass
for the muon and the second neutrino eigenstate, and secondly, it generates
a large atmospheric neutrino mixing. The squared mass matrices $H_{\nu }=$ $%
M_{\nu }M_{\nu }^{\dagger }$ and $H_{l}=$ $M_{l}M_{l}^{\dagger }$ are now
significantly different: 
\begin{equation}
\begin{array}{lll}
\begin{array}{c}
H_{\nu }=\rho _{\nu }\ D_{L}\cdot \Gamma _{\nu }\cdot D_{L} \\ 
\\ 
H_{l}=\rho _{l}\ D_{L}\cdot \Gamma _{l}\cdot D_{L}%
\end{array}
& \ ;\ \ \ \  & \Gamma _{\nu ,l}=\left[ 
\begin{array}{ccc}
1 & 1 & -c_{2\theta } \\ 
1 & 1 & -c_{2\theta } \\ 
-c_{2\theta } & -c_{2\theta } & 1%
\end{array}
\right] _{\nu ,l}%
\end{array}
\label{hnch1}
\end{equation}
with $c_{2\theta _{\nu ,l}}\equiv \cos (2\theta _{\nu ,l})$ and 
\begin{equation}
\begin{array}{ccc}
\cos (\theta _{\nu })=\frac{b_{\nu _{3}}}{\rho _{\nu }} & \ ;\ \ \ \  & \cos
(\theta _{l})=\frac{b_{l_{3}}}{\rho _{l}}%
\end{array}
\label{tanbeta}
\end{equation}
Using a similar parameterization, as for $\rho _{\nu ,l}$ in (\ref{hnch}),
for $\rho _{L}=\sqrt{a_{L1}^{2}+a_{L2}^{2}+a_{L2}^{2}}$, one finds, in this
approximation, the following squared roots of eigenvalues of $H_{l}$ and $%
H_{\nu }$ respectively: 
\begin{equation}
\begin{array}{lll}
m_{e}\equiv 0 & \ ;\ \ \ \  & m_{\nu _{1}}\equiv 0 \\ 
&  &  \\ 
m_{\mu }=\frac{\rho _{L}\rho _{l}}{\sqrt{2}}\sqrt{1-\sqrt{1-s_{2\theta
_{L}}^{2}s_{2\theta _{l}}^{2}}} & \ ;\ \ \ \  & m_{\nu _{2}}=\frac{\rho
_{L}\rho _{\nu }}{\sqrt{2}}\sqrt{1-\sqrt{1-s_{2\theta _{L}}^{2}s_{2\theta
_{\nu }}^{2}}} \\ 
&  &  \\ 
m_{\tau }=\frac{\rho _{L}\rho _{l}}{\sqrt{2}}\sqrt{1+\sqrt{1-s_{2\theta
_{L}}^{2}s_{2\theta _{l}}^{2}}} & \ ;\ \ \ \  & m_{\nu _{3}}=\frac{\rho
_{L}\rho _{\nu }}{\sqrt{2}}\sqrt{1+\sqrt{1-s_{2\theta _{L}}^{2}s_{2\theta
_{\nu }}^{2}}}%
\end{array}
\label{masses}
\end{equation}
where $s_{2\theta _{\nu ,l}}\equiv \sin (2\theta _{\nu ,l})$ and $s_{2\theta
_{L}}\equiv \sin (2\theta _{L})$, with a suitable parametrization for $%
(a_{1},a_{2},a_{3})_{L}$ $=$ $\rho _{L}(\sin (\varphi _{L})\sin (\theta
_{L}) $, $\cos (\varphi _{L})\sin (\theta _{L})$, $\cos (\theta _{L}))$. In
addition, the matrices $O_{\nu ,l}$ which diagonalize $H_{\nu ,\,l}$ in (\ref%
{hnch1}), will have the following form 
\begin{equation}
O_{\nu ,l}=O_{\varphi _{L}}\cdot (O_{\overline{\theta }_{\nu ,l}})^{T}
\label{odiag}
\end{equation}
where 
\begin{equation}
\begin{array}{ccc}
O_{\varphi _{L}}=\left[ 
\begin{array}{ccc}
\cos (\varphi _{L}) & \sin (\varphi _{L}) & 0 \\ 
-\sin (\varphi _{L}) & \cos (\varphi _{L}) & 0 \\ 
0 & 0 & 1%
\end{array}
\right] & \ ;\ \ \ \  & O_{\overline{\theta }_{\nu ,l}}=\left[ 
\begin{array}{ccc}
1 & 0 & 0 \\ 
0 & \cos (\overline{\theta }) & \sin (\overline{\theta }) \\ 
0 & -\sin (\overline{\theta }) & \cos (\overline{\theta })%
\end{array}
\right] _{\nu ,l}%
\end{array}
\label{odiag1}
\end{equation}
with $\tan (2\overline{\theta }_{\nu ,l})=\tan (2\theta _{L})\cos (2\theta
_{\nu ,l})$. Notice that $O_{\varphi _{L}}$, containing the angle $\varphi
_{L}$ from the parameterization of $(a_{1,}a_{2},a_{3})_{L}$, appears both
in $O_{\nu }$ and $O_{l}$, and thus 
\begin{equation}
U_{MNS}=O_{\overline{\theta }_{l}}\cdot (O_{\overline{\theta }_{\nu }})^{T}=%
\left[ 
\begin{array}{ccc}
1 & 0 & 0 \\ 
0 & \cos (\overline{\theta }) & \sin (\overline{\theta }) \\ 
0 & -\sin (\overline{\theta }) & \cos (\overline{\theta })%
\end{array}
\right] \ ;\ \ \ \overline{\theta }=\overline{\theta }_{l}-\overline{\theta }%
_{\nu }  \label{nosolar}
\end{equation}
One may have large atmospheric mixing, but, unfortunately, no solar mixing%
\footnote{%
Again, small variations in $\kappa _{ij}$ will not be sufficient to generate
large solar angles.}. Therefore, in order to obtain sufficient large solar
mixing we must have $(-)$ signs at different places in the mass matrices of
the neutrinos and charged leptons. \newline
\newline
$\bullet $ Next, we study the case $\kappa _{33}^{\nu }=\kappa _{13}^{l}=-1$
with all the other $\kappa _{ij}^{\nu }=\kappa _{ij}^{l}=1$. Due to the
different position of the $(-)$ sign in the charged lepton mass matrix, we
will obtain diagonalizing matrices which are significantly different. Then,
it is possible to have large atmospheric and solar mixings. The permutation,
induced by the different position of the $(-)$ sign, results in the
following diagonalizing matrix for the charged leptons: 
\begin{equation}
\begin{array}{lll}
O_{l}=P\cdot O_{\widehat{\varphi }_{L}}\cdot (O_{\overline{\theta }_{l}})^{T}
& \ ;\ \ \ \  & P=\left[ 
\begin{array}{ccc}
0 & 0 & 1 \\ 
0 & 1 & 0 \\ 
1 & 0 & 0%
\end{array}
\right]%
\end{array}
\label{ochl}
\end{equation}
where $O_{\widehat{\varphi }_{L}}$ is similar to $O_{\varphi _{L}}$, i.e. a
rotation of the first and second coordinates, but where the angle $\widehat{%
\varphi }_{L}$ comes from a different parameterization: $%
(a_{1,}a_{2},a_{3})_{L}=\rho _{L}(\cos (\widehat{\theta }_{L})$, $\cos (%
\widehat{\varphi }_{L})\sin (\widehat{\theta }_{L})$, $\sin (\widehat{%
\varphi }_{L})\sin (\widehat{\theta }_{L}))$, as a result of the
permutation. As in (\ref{odiag1}), one has $\tan (2\overline{\theta }%
_{l})=\tan (2\widehat{\theta }_{L})\cos (2\theta _{l})$: now related to the
new $\widehat{\theta }_{L}$ of this parameterization. It is clear that, due
to the permutation, $O_{\varphi _{L}}$ and $O_{\widehat{\varphi }_{L}}$ are
not equal. In addition, the product $O_{\widehat{\varphi }_{L}}^{T}\cdot
P\cdot O_{\varphi _{L}}$, appearing in $U_{MNS}$, does not cancel, and
therefore, one may have large mixing.

\end{document}